\documentclass[12pt,twoside]{article}
\usepackage{xrb2007}
\usepackage{graphicx}

\newcommand{\nb}{Nuclear Bulge}

\newcommand{\nir}{near-infrared}

\begin{document}

% select your session by uncommenting the appropriate line
%\session{Jets}
%\session{Jet and Black Hole Binaries}
%\session{Faint Galactic XRB Populations}
%\session{Faint XRBs and Galactic LMXBs}
\session{Obscured XRBs and INTEGRAL Sources}
%\session{ULXs}
%\session{Extragalactic Populations}
%\session{Future Missions and Surveys}
%\session{Population Synthesis}

\shortauthor{Gosling et al.}
\shorttitle{The Nuclear Bulge extinction}

\title{The Nuclear Bulge extinction} \author{Andrew J. Gosling,
Katherine M. Blundell} \affil{Oxford Astrophysics, Denys Wilkinson
Building, Keble Road, Oxford, OX1 3RH}

\author{Reba M. Bandyopadhyay} \affil{Department of Astronomy,
University of Florida, 210 Bryant Space Science Center, Gainesville,
FL 32611-2055} 

\author{Phil Lucas} \affil{School of Astronomy, University of
Hertfordshire, College Lane, Hatfield, Herts AL10 9AB}

\begin{abstract}
We present a new, high resolution ($5\arcsec$ per pixel) \nir\
extinction map of the \nb\ using data from the UKIDSS-GPS. Using
photometry from the $J$, $H$ and $K$-bands we show that the extinction
law parameter $\alpha$ is also highly variable in this region on
similar scales to the absolute extinction. We show that only when this
extinction law variation is taken into account can the extinction be
measured consistently at different wavelengths.
\end{abstract}

\section{Introduction}

Study of the \nb\ (NB) is extremely difficult as it is one of the most
highly obscured regions of the Galaxy. The extinction towards the NB
is so great that it is almost impossible to observe at visual
wavelengths. It is necessary to observe in the \nir\ (NIR) to obtain
imaging similar to that normally obtained in visual. Previous measures
of the NIR extinction towards the NB \citep{catc90, schu99, dutr03}
found the extinction to be highly spatially variable. We present a new
mapping of the extinction towards the NB at a much higher resolution
than previously possible. 

\section{Extinction mapping}

We mapped the extinction across a region of approximately $2\deg
\times 2\deg$ centred on Sgr A* using near-infrared data from the
(United Kingdom Infrared Deep Sky Survey - Galactic Plane Survey)
UKIDSS-GPS \citep{luca07}. Using a 2:1 over-sampling, we obtained a
resolution of 5\arcsec\ over the entire region, an order of magnitude
better than previous maps. At each position on the map we use the
median colour-indices of the stars within a 10\arcsec\ box to
calculate the local extinction law parameter $\alpha$ using the
equation:
%\begin{equation} \label{eq:extlaw}
\begin{center}
$\frac{\left< \lambda_1 - \lambda_2 \right>}{\left< \lambda_2 -
\lambda_3 \right>} = \frac{\left( \frac { \lambda_{2} }{ \lambda_{1} }
\right)^{\alpha} - 1}{1 - \left( \frac { \lambda_{2} }{ \lambda_{3} }
\right)^{\alpha} }$
\end{center}
%\end{equation}
\noindent
 Until recently, this extinction law parameter which indicates the
relationship between extinction and wavelength as $A_{\lambda} \propto
\lambda^{-\alpha}$ was thought to be a constant $\alpha \sim 2$
\citep{mart90, riek85, nish06} for the NB. We show that it is in fact
highly variable in the NB. Using a local extinction law calculation,
we then measure the absolute extinction in the $J$, $H$, and
$K$-bands using the equations:
%\begin{equation} \label{eq:ext1}
\begin{center}
$A_{\lambda_1} = \frac{\left< \lambda_1 - \lambda_2 \right>}{1 - \left(
\frac { \lambda_{1} }{ \lambda_{2} } \right)^{\alpha}}
%\end{equation}
%\begin{equation} \label{eq:ext2}
\hspace{20mm}
A_{\lambda_2} = \frac{\left< \lambda_1 - \lambda_2 \right>}{\left(
\frac { \lambda_{2} }{ \lambda_{1} } \right)^{\alpha} - 1}$
\end{center}
%\end{equation}

Our final calculation of absolute extinction takes into account the
variation in the extinction law as well as variation in the amount of
absolute extinction at each location. Using data from the 3 bands,
$J$, $H$, and $K$, we are able to calculate three measures of the
extinction law and two measures of the absolute extinction for each
band at each position in our map. Full details of this method will be
presented in Gosling et al. (2008).

\section{Measured extinction}

\begin{figure}
  \begin{center}
    \includegraphics[width=0.4\textwidth, height=0.2\textwidth]{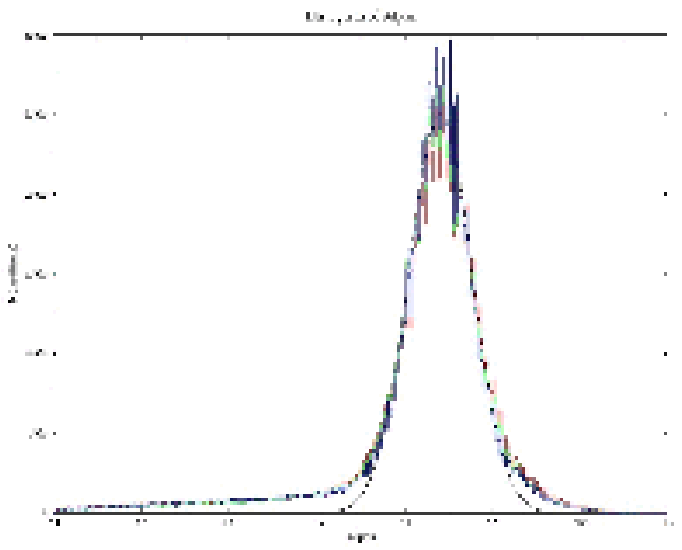}
    \includegraphics[width=0.4\textwidth, height=0.2\textwidth]{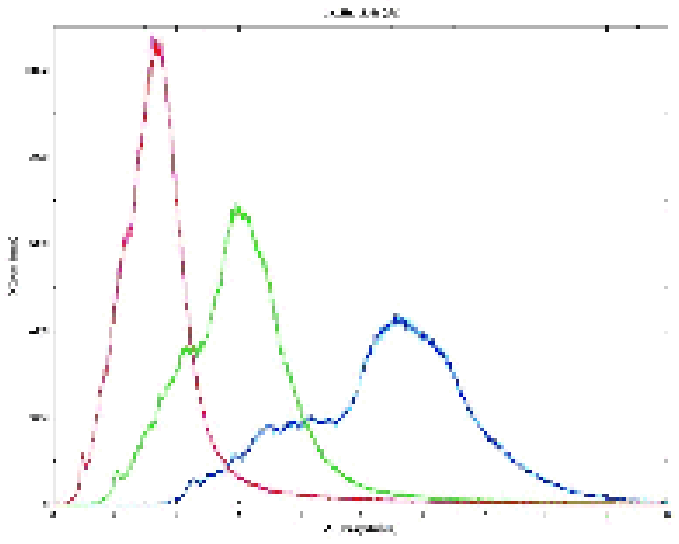}
    \caption[]{Histograms of the values of extinction law parameter
      $\alpha$ measured across the NB ({\it left} figure) and the
      absolute extinction in magnitudes in the $J$, $H$ and $K$-bands
      ({\it blue}, {\it green} and {\it red} respectively in the {\it
      right} figure). The fit to the measured $\alpha$ is shown as a
      gaussian with mean of 2.20 and standard deviation of 0.18.}
    \label{f:graphs}
  \end{center}
\end{figure}

Figure \ref{f:graphs} shows histograms of the value of the extinction
law parameter $\alpha$ and the values of absolute extinction measured
from the three colour-indices ratios.

The same distribution of $\alpha$ is measured from all three colour
indices ratios, and this distribution can be described by a gaussian
with mean 2.20 and standard deviation 0.18. We suggest that a specific
extinction law calculation be undertaken for all observations in the
NB if possible, however, when this is not possible, we recommend a
value of $\alpha = 2.2 \pm 0.54$ be used. In such cases the possible
variation from this mean of $\alpha$ should be considered as it can
have a large effect on the relative values of absolute extinction
calculated.

\begin{figure}
  \begin{center}
    \includegraphics[width=0.5\textwidth]{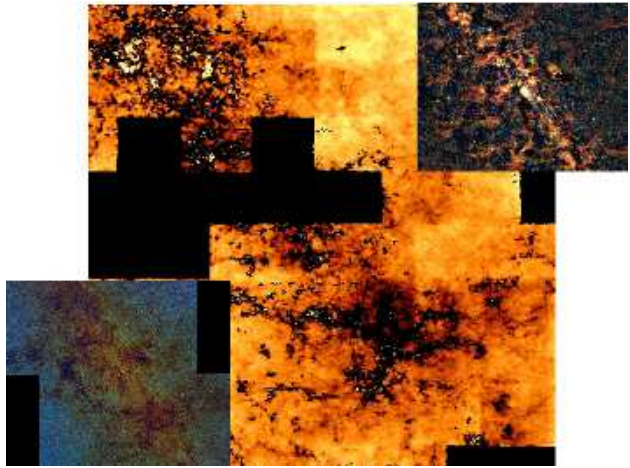}
    \caption[]{{\it Main Figure}: $A_K$ map of the Nuclear Bulge made
      using data from the UKIDSS-GPS survey (see text) with resolution
      of $5\arcsec$. The colour scale ranges from 0 (white) to 3.5
      (black) magnitudes of extinction. Square black regions are area
      for which there is no data as yet in the UKIDSS catalogue. There
      is turnover to low extinction within some regions of high
      extinction where only foreground stars are observed as all
      background NB stars are obscured. {\it Bottom-left}: $J$, $H$,
      and $K$-band false-colour image of the NB from the UKIDSS-GPS,
      the photometry of which was used to produce the extinction map.
      {\it Top-right}: {\it Spitzer} false-colour image of the same
      region using the 3.6, 5.4 and 8.0 \micron\ channels scaled to
      reveal dust emission. Note the correspondence between the
      emission in this image and the regions of high extinction from
      the UKIDSS map.}
    \label{f:extmap}
  \end{center}
\end{figure}

Using the specific, local extinction law, we calculated the value of
absolute extinction in the $J$, $H$ and $K$-bands from all three
colour indices ratios. We measured the extinction in magnitudes to be
$1.8 < A_J < 10.0$, $0.7 < A_H < 7.0$ and $0.3 < A_K < 6.0$. All three
distributions have high extinction tails extending to higher values of
absolute extinction than these ranges. Figure \ref{f:extmap} shows the
resultant extinction map as well as a false colour NIR and
mid-infrared image of the same region for comparison.

We repeated the extinction calculation described above using a single
extinction law. We used a $\chi^2$-test comparison between the
extinction calculated for different bands to show that the calculation
using the variable extinction law produced statistically consistent
extinction in all three bands (differences within 3$\sigma$ of the
overall distribution), whereas using a single extinction value
produced statistically inconsistent extinction between bands
(differences of 10-50$\sigma$ of the overall distribution).

\section{Conclusions}

We have presented a new, highly detailed map of the NIR extinction
towards the NB using data from the UKIDSS-GPS. We find both
variation in the degree of absolute extinction, and in the extinction
law parameter $\alpha$. We show that the extinction law is not
``universal'' as had previously been thought \citep{mart90, riek85},
but is highly variable, and that only when this variation is taken
into account can absolute extinction be calculated consistently at
different wavelengths. We recommend that this variable extinction law
be taken into account in all future extinction corrections. We intend
to compare our extinction law and absolute extinction maps to dust
emission observed with {\it Spitzer} ({\it top-right} insert in Figure
\ref{f:extmap}) to better determine the properties of the extincting
material and so the distribution of gas and dust in the Bulge.

\end{document}